\documentclass[aps,prb,showpacs,twocolumn]{revtex4}
\usepackage{amsmath,amssymb,bbm,mathrsfs}
\usepackage{graphicx}
\usepackage{epsfig}
\newcommand{\al}{\alpha}
\newcommand{\be}{\beta}
\newcommand{\g}{\gamma}
\newcommand{\de}{\delta}
\newcommand{\e}{\epsilon}
\newcommand{\z}{\zeta}

\newcommand{\la}{\lambda}
\newcommand{\La}{\Lambda}
\newcommand{\mi}{\mu}

\newcommand{\p}{\pi}
\newcommand{\ro}{\rho}
\newcommand{\s}{\sigma}

\newcommand{\Y}{\Upsilon}

\newcommand{\w}{\omega}

\newcommand{\De}{\Delta}
\newcommand{\G}{\Gamma}
\renewcommand{\S}{\Sigma}
\newcommand{\D}{\Delta}
\newcommand{\ta}{\tau}
\newcommand{\E}{\Xi}
\newcommand{\pd}{\partial}

\newcommand{\round}[1]{\left({#1}\right)}
\renewcommand{\square}[1]{\left[{#1}\right]}
\newcommand{\abs}[1]{\left|{#1}\right|}
\newcommand{\cvec}[2]{\round{\begin{array}{c}#1\\#2\end{array}}}
\newcommand{\cvecf}[4]{\round{\begin{array}{c}#1\\#2\\#3\\#4\end{array}}}
\newcommand{\cvece}[8]{\round{\begin{array}{c}#1\\#2\\#3\\#4\\#5\\#6\\#7\\#8
\end{array}}}
\newcommand{\mat}[4]{\round{\begin{array}{cc}#1&#2\\#3&#4\end{array}}}

\newcommand{\rvec}[2]{\round{\begin{array}{cc}#1&#2\end{array}}}

\newcommand{\ang}[1]{\left\langle #1\right\rangle}
\newcommand{\dang}[1]{\left\langle\left\langle #1\right\rangle\right\rangle}

\newcommand{\beq}{\begin{equation}}
\newcommand{\eeq}{\end{equation}}
\newcommand{\Beq}{\begin{eqnarray}}
\newcommand{\Eeq}{\end{eqnarray}}
\newcommand{\beqm}{\begin{multline}}
\newcommand{\eeqm}{\end{multline}}
\newcommand{\tri}{\triangle}
\newcommand{\bsp}{\begin{split}}
\newcommand{\esp}{\end{split}}
\newcommand{\down}{\downarrow}
\newcommand{\up}{\uparrow}

\begin{document}

\title{Nonequilbrium-induced metal-superconductor quantum phase transition in graphene}
\author{So Takei$^1$ and Yong Baek Kim$^{1,2}$}
\affiliation{$^{1}$Department of Physics, The University of Toronto, Toronto,
Ontario M5S 1A7, Canada\\
$^2$School of Physics, Korea Institute for Advanced Study, Seoul 130-722, Korea}
\date{\today}

\begin{abstract}
We study the effects of dissipation and time-independent nonequilibrium drive on an open superconducting
graphene. In particular, we investigate how dissipation and nonequilibrium effects 
modify the semi-metal-BCS quantum phase transition that occurs at half-filling in equilibrium graphene 
with attractive interactions. 
Our system consists of a graphene sheet sandwiched by two semi-infinite three-dimensional Fermi liquid reservoirs, 
which act both as a particle pump/sink and a source of decoherence. A steady-state charge current is established 
in the system by equilibrating the two reservoirs at different, but constant, chemical potentials.
The graphene sheet is described using the attractive Hubbard model in which the interaction is decoupled
in the $s$-wave channel. 
The nonequilibrium BCS superconductivity in graphene is formulated using the Keldysh path integral formalism, 
and we obtain generalized gap and number density equations valid for both zero and finite voltages. The 
behaviour of the gap is discussed as a function of both attractive interaction strength and electron densities 
for various graphene-reservoir couplings and voltages. We discuss how tracing out the dissipative environment 
(with or without voltage) leads to decoherence of Cooper pairs in the graphene sheet, 
hence to a general suppression of the gap order parameter at all densities. For weak enough
attractive interactions we show that the gap vanishes even for electron densities away from half-filling,
and illustrate the possibility of a dissipation-induced metal-superconductor quantum phase transition. 
We find that the application of small voltages does not alter the essential features of the gap as
compared to the case when the system is subject to dissipation alone (i.e. zero voltage). 
The possibility of tuning the system through the metal-superconductor quantum critical point using 
voltage is presented. 
\end{abstract}
\maketitle

\section{Introduction}
The landmark experimental realization of an isolated graphite monolayer, or 
graphene\cite{novoselov1,novoselov2}, has sparked intense theoretical and
experimental interest in the material over the last few years\cite{rise}. 
A source of interest in 
the study of graphene is the unique properties of its charge carriers. At
low energies, these charge carriers mimic relativistic particles, and are  
most naturally described by the (2+1)-dimensional Dirac equation with an
effective speed of light, $c\sim v_F\sim 10^6 \mbox{ms}^{-1}$. The fact that
graphene is an excellent condensed-matter analogue of (2+1)-dimensional 
quantum electrodynamics (QED) has been known to theorists  
for over 20 years\cite{semenoff,fradkin,haldane}. 
However, it was not until the spectacular experimental realization of isolated 
graphene that experimentalists began observing signatures of the QED-like 
spectrum in their laboratories. Consequences of graphene's
unique electronic properties have been revealed in the context of anomalous integer 
quantum-Hall effect\cite{novoselov3,qhe} and minimum quantum conductivity in 
the limit of vanishing carrier concentrations\cite{novoselov3}.

In addition to its importance in fundamental physics, graphene is expected to make 
a significant impact in the world of nano-scale electronics. Research efforts in developing
graphene-based electronics have been fueled by a strong anticipation that 
it may supplement the silicon-based technology which is nearing its limits\cite{rise}. 
Graphene is a promising material for future nanoelectronics because
of its exceptional carrier mobility which remains robustly high for a large range
of temperatures, electric-field-induced concentrations\cite{novoselov1,novoselov2,
novoselov3,qhe}, and chemical doping\cite{mobility}. Indeed, recent experiments
have explored the possibilities of in-plane graphene heterostructures by engineering arbitrary 
spatial density variation using local gates\cite{pkim,lemme,goldg}. The application
of local gate techniques to graphene marks an important first step on the road 
towards graphene-based electronics.

From a theoretical point of view realizing graphene nanoelectronics requires a theoretical
understanding of open nonequilibrium graphene. Naturally, graphene in nano-circuits 
is subject to decoherence effects due to its coupling to external leads via tunnel junctions. 
Furthermore, a nonequilibrium treatment of graphene becomes necessary when a charge 
current is driven through it. To this date, effects of dissipation and nonequilibrium drive on 
graphene electronic properties have not been addressed. The focus of this paper is to show a
theoretical framework in which these effects can be studied and illustrate how they give rise
to striking influences on the equilibrium properties of graphene.

This work considers dissipation and nonequilibrium effects on superconducting graphene. 
Besides the possibility of superconductivity in graphene by proximity 
effect\cite{heersche}, some works suggested the potential of achieving plasmon-mediated singlet 
superconductivity in graphene\cite{castro,qp}. 
Several groups have investigated the equilibrium mean-field theory of superconductivity in 
graphene using the attractive Hubbard model on the honeycomb lattice. 
Uchoa and Castro-Neto\cite{castro} studied spin singlet superconductivity in graphene at various fillings
by considering both the usual $s$-wave pairing as well as pairing with $p+ip$ orbital symmetry 
permitted by graphene's honeycomb lattice structure. Zhao and Paramekanti\cite{erhai}
examined the possibility of $s$-wave superconductivity on the honeycomb lattice.
Both works show that (in the absence of $p$-wave pairing) half-filled graphene displays a 
semimetal-superconductor quantum critical point at a finite critical attractive interaction strength 
$u_c$. Away from half-filling, the system exhibits Cooper instability at any finite
$u$ and thus undergoes the usual BCS-BEC crossover as $u$ is increased.
The difficulty in achieving superconductivity at half-filling is a result of the vanishing 
density of states at the Dirac point and the absence of electron screening. 

In this work, the superconducting graphene sheet is subjected to dissipation and nonequilibrium drive
by coupling it to two semi-infinite particle reservoirs via tunnel junctions. The geometry of the
system is shown in Fig.\ref{fig:system}. While the two reservoirs are independently held in 
thermal and chemical equilibrium at all times, an out-of-plane steady-state current through 
graphene is established by equilibrating the reservoirs at two different, but constant, 
chemical potentials. The leads act as infinite reservoirs and are 
assumed to be held at a common temperature $T$ at all times.
Nonequilibrium theory of BCS superconductivity is formulated
using the Keldysh path integral formalism, and the resulting nonequilibrium mean-field
equations are used to investigate the gap behaviour at and near half-filling
for various attractive interaction strengths. The gap is plotted in the parameter space of
filling $n$ and the interaction strength $u$ (see Fig.\ref{fig:gap}), and our results can 
be directly compared to the gap phase diagram in Fig.2 of the work by Zhao 
and Paramekanti\cite{erhai}.

Our main results are now qualitatively summarized. We find that the gap is generally suppressed in 
the presence of leads. As the paper will discuss in detail, the key to understanding our findings is to 
notice that the dissipation of electrons into the leads act as a pair-breaking mechanism for the Cooper
pairs in the central graphene sheet. This mechanism, and hence the suppression, is present at both 
zero and finite voltages and for all electron densities. As a consequence, the Fermi liquid ground state
of the system remains stable against Cooper pairing up to some density-dependent finite attractive 
interaction strength $u_c(n)$ at all densities. With respect to the gap phase diagram, dissipation 
gives rise to a finite region around half-filling in which the gap vanishes (see Fig.\ref{fig:gap}). 
From these results, we infer that dissipation induces a metal-superconductor quantum phase transition 
at all fillings, for which the tuning parameter is the attractive interaction strength $u$. The 
qualitative behaviour of the gap is not appreciably different in the zero and finite voltage cases as 
long as the voltage is small, i.e. $V\ll\G$, where $\G$ denotes the average tunneling rate of electrons 
between graphene and the two leads. Finite voltage modifications, however, result due to 
voltage-induced changes in the graphene electron density.

The paper is organized as follows. In Sec.\ref{model}, we introduce the Hamiltonian which models
our heterostructure. The mean-field treatment of the model is formulated on the Keldysh 
contour in Sec.\ref{keldysh}. In Sec.\ref{mfeqs}, the nonequilibrium gap and number density equations
will be derived. The results are presented in Sec.\ref{results}. The effects of dissipation in the absence of
voltage is discussed in Sec.\ref{results0v} while the finite voltage effects are included in Sec.\ref{resultsfv}.
We conclude in Sec.\ref{conclusion}.

\section{Model}
\label{model}
\begin{figure}[h]
\begin{center}
\includegraphics[scale=0.8]{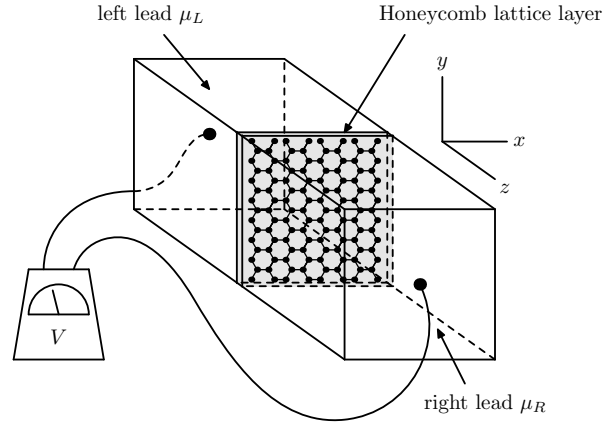}
\caption{\label{fig:system} A schematic of the system considered. Chemical potential mismatch
in the two leads will lead to a charge current parallel to the $z$-axis.}
\end{center}
\end{figure}

The lead-graphene-lead heterostructure considered in this work is shown in Fig.\ref{fig:system}. Graphene 
is located on the $z=0$ plane, and each of its sites is labeled using two coordinates 
${\bf r}_i=(x_i,y_i,z_i\equiv 0)$. 
The semi-infinite metallic leads extend from both sides of the graphene sheet for $z>0$ and $z<0$. 
We assume the leads are separated from graphene by thin insulating barriers, and the tunneling of 
electrons through each of the barriers can be described by phenomenological tunneling parameters. 
Full translational symmetry is present along the planes parallel to the $xy$-plane for $z\ne0$ while only the 
discrete translational symmetry of the graphene lattice is present at $z=0$. The leads are 
assumed to be in thermal equilibrium with their continuum of states occupied according to the Fermi-Dirac 
distribution, $f_\al(\w)=\square{1+\mbox{exp}\round{\be(\w-\mi_\al)}}^{-1}$, where 
$\al=L(\mbox{left}),R(\mbox{right})$ labels the leads. An electric potential bias is set up in the 
out-of-plane direction by tuning the chemical potentials 
of the leads to different values. 

The Hamiltonian consists of three parts,
\beq
H=H_{sys}+H_{res}+H_{sys-res}.
\eeq
The central graphene sheet is modeled using the attractive Hubbard model on the honeycomb lattice. 
The kinetic term is a tight-binding description for the $\p$-orbitals of carbon that includes nearest- and 
next-nearest-neighbour hopping processes. The on-site interaction strength is parametrised by $U$.
The Hamiltonian for the layer is 
\beqm
H_{sys}=-t\sum_{\ang{i,j},\s}(c^\dag_{i,\s}c_{j,\s}+h.c.)\\
-t'\sum_{\dang{i,j},\s}(c^\dag_{i,\s}c_{j,\s}+h.c.)
-U\sum_{i}c^\dag_{i,\up}c^\dag_{i,\down}c_{i,\down}c_{i,\up}.
\label{hsys}
\end{multline}
$c^\dag_{i,\s}$ ($c_{i,\s}$) creates (annihilates) electrons on site ${\bf r}_i$ of
the graphene honeycomb lattice with spin $\s$ ($\s=\up,\down$). 
$U$ is assumed positive due to attractive interaction, and $t$ and $t'$ are the 
nearest- and next-nearest neighbour hopping parameters, respectively. Specific values 
for $t$ and $t'$ have been estimated\cite{reichetal} by comparing a tight-binding description
to first-principle calculations. Following their estimates, we take $t=2.7$eV 
and fix $t'/t=0.04$. 

\begin{figure}[h]
\begin{center}
\includegraphics[scale=0.8]{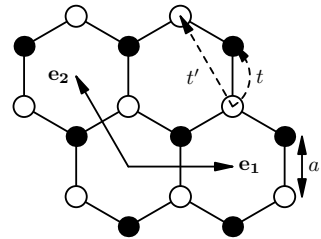}
\caption{\label{fig:lattice} Graphene honeycomb lattice. ${\bf e_1}$ and ${\bf e_2}$ are the 
unit-cell basis vectors of graphene with lattice constant $\sqrt{3}a\approx 2.46\mathring{A}$ 
($a\approx 1.42\mathring{A}$). A unit cell contains two carbon atoms belonging to the two 
sublattices $A$ (white circles) and $B$ (black circles). All nearest- and next-nearest-neighbour 
hopping matrix elements are $-t$ and $-t'$, respectively.}
\end{center}
\end{figure}

The honeycomb lattice can be described in terms of two inter-penetrating triangular
sublattices, $A$ and $B$ (see Fig.\ref{fig:lattice}). Each unit cell is composed of
two atoms, one each of type $A$ and type $B$. Primitive translation vectors, 
${\bf e_1}$ and ${\bf e_2}$, are
\beq
{\bf e}_1=(\sqrt{3},0)\quad {\bf e}_2=(-\sqrt{3}/2,3/2)\quad 
{\bf e}_3={\bf e}_1+{\bf e}_2,
\eeq
where they are expressed in units of $a$, the distance between two nearest carbon atoms.
Any $A$ atom is connected to its nearest neighbours on the $B$ lattice by three
vectors
\Beq
{\bf d}_1&=&(0,1)\nonumber\\
{\bf d}_2&=&(-\sqrt{3}/2,-1/2)\\
{\bf d}_3&=&(\sqrt{3}/2,-1/2).\nonumber
\Eeq
In momentum space, the kinetic term reads 
\beq
H^K_{sys}={1\over N_\tri}\sum_{{\bf k},\s}\rvec{a^\dag_{{\bf k},\s}}
{b^\dag_{{\bf k},\s}}\mat{\la_{\bf k}}{g^*_{\bf k}}{g_{\bf k}}{\la_{\bf k}}
\cvec{a_{{\bf k},\s}}{b_{{\bf k},\s}},
\label{hkms}
\eeq
where 
\Beq 
\la_{\bf k}&=&-t'\round{\sum_{i=1}^3e^{i{\bf k}\cdot{\bf e}_i}+c.c}\\
g_{\bf k}&=&-t\sum_{i=1}^3e^{i{\bf k}\cdot{\bf d}_i}.
\Eeq
Components of the pseudospinor, $a^\dag_{{\bf k},\s}$ and $b^\dag_{{\bf k},\s}$, describe 
quasiparticles that belong to sublattice $A$ and $B$, respectively. Here, $N_\tri$ denotes the number 
of lattice sites in a triangular sublattice. $N=2N_\tri$ will denote the total number of sites on the honeycomb 
lattice. 

\begin{figure}[h]
\begin{center}
\includegraphics[scale=0.8]{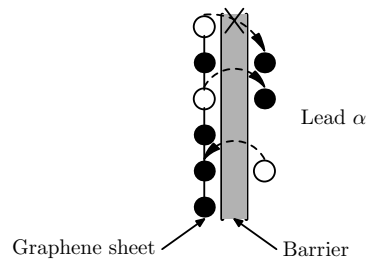}
\caption{\label{fig:tunnel} A diagram illustrating the type of tunneling process that are considered
in this work. The diagram is an edge-on view of the interface between the graphene sheet and a lead. 
The only tunneling events that are allowed are those in which the ($x$,$y$) coordinates of electrons remain 
unaltered. Thus, while the lower two processes in the diagram are allowed, tunneling of the type shown at the 
top is disallowed.}
\end{center}
\end{figure}

Coupling between leads and the graphene sheet is modeled using the following Hamiltonian,
\beqm
H_{sys-res}=\\
\int{dk_z\over 2\p}\sum_{\al=L,R}\sum_{i,\s}\z_{\al}\round{C^\dag_{i,\s,\al,k_z}
c_{i,\s}+h.c.}.
\label{hsr}
\end{multline}
$\z_{\al}$ is a phenomenological tunneling matrix that describes the tunneling of an electron 
between site $i$ on the graphene sheet and an adjacent site on lead $\al$ (see Fig.\ref{fig:tunnel}). 
We only consider lead-graphene tunneling processes in which $(x,y)$ coordinates of the electron 
in the initial and final states are the same. This assumption simplifies various computational steps
without altering the qualitative features of the final results. 
$C^\dag_{i,\s,\al,k_z}$ creates an electron in lead $\al$ 
at coordinates $(x_i,y_i)$ with spin $\s$ and longitudinal momentum $k_z$. We assume here that 
the tunneling parameters are independent of frequency and momentum but maintain their lead 
dependence in order to describe possible asymmetries in the lead-layer couplings. 
In momentum space, the tunneling Hamiltonian 
Eq.\ref{hsr} becomes
\beqm
H_{sys-res}=\sum_\al \z_\al\int{dk_z\over 2\p}{1\over N_\triangle}\sum_{{\bf k},\s}\\
\round{A^{\dag}_{{\bf k},k_z,\s,\al}a_{{\bf k},\s}+
B^{\dag}_{{\bf k},k_z,\s,\al}b_{{\bf k},\s}+h.c.}.
\label{hsrm}
\end{multline}
The inplane momentum, ${\bf k}$, is the component of momentum parallel to the graphene 
plane and the out-of-plane momentum, $k_z$, is its component normal to the plane. 
$A^{\dag}_{{\bf k},k_z,\s,\al}$ ($B^{\dag}_{{\bf k},k_z,\s,\al}$) corresponds to an electron mode 
propagating in "sublattice $A(B)$" in lead $\al$ with spin $\s$ and wavevector ${\bf k}$.
Although the full inplane translational symmetry of the leads implies that ${\bf k}$ can take 
on any value in $\mathbb{R}^2$ the tunneling assumption 
(see Fig.\ref{fig:tunnel}) tells us that the only modes that tunnel are those with ${\bf k}$ values 
that are the allowed modes of the triangular sublattices in the graphene sheet. 
All other inconsequential modes can eventually be integrated out in the path integral sense and 
will merely contribute a multiplicative factor in front of the partition function. Therefore, we will 
not consider these modes further.

Both leads are assumed to be Fermi liquids
\beqm
H_{res}=\sum_{\al,\La}{1\over N_\tri}\int{dk_z\over 2\p}\sum_{{\bf k},k_z,\s}\\
\e_{{\bf k},k_z}\round{A^{\dag}_{{\bf k},k_z,\s,\al}A_{{\bf k},k_z,\s,\al}
+B^{\dag}_{{\bf k},k_z,\s,\al}B_{{\bf k},k_z,\s,\al}},
\label{hleads}
\end{multline}
with a separable dispersion
\beq
\e_{{\bf k},k_z}=\e_{\bf k}+\e_{k_z}=\frac{\abs{\bf k}^2}{2m_e}+\frac{k_z^2}{2m_e}.
\eeq
Besides their role as a particle pump/sink, the leads play an important role as
a heat sink. An important assumption we make is that any heat generated in the
interacting region due to the application of a transverse electric field is efficiently 
dissipated into the leads so as to prevent build up of heat in the region. This
is a well-justified assumption because the leads are assumed to be infinite and the
interacting region has a thin profile. 

In equilibrium ($\mi_{res}=\mi_R=\mi_L$), the central system is expected to reach chemical 
equilibrium with the reservoirs in the long-time limit so that $\mi_{sys}=\mi_{res}$. In the
out-of-equilibrium case, the system is coupled to two reservoirs that are not in chemical
equilibrium. Therefore, although the electron distribution in the interacting system reaches
a static form in the long-time limit, it is in no way expected to have an equilibrium form 
due to constant influx (outflux) of particles from (into) the leads. 

\section{Keldysh path integral formulation}
\label{keldysh}
In this section, we formulate a theory of nonequilibrium BCS superconductivity in graphene 
using the Keldysh functional integral formalism. The theory is first expressed in terms of a
Keldysh partition function using coherent states of fields defined on the time-loop Keldysh contour, 
$\mathcal{C}$. Following a Hubbard-Stratonovic decoupling of the quartic interaction term in the pair channel
a BCS theory for superconducting graphene is obtained by assuming a 
static, homogeneous gap, integrating out both leads and graphene electrons, and extremizing the 
effective action with respect to the gap. The resulting mean-field equations, which are a nonequilibrium 
generalization of the corresponding equilibrium equations\cite{erhai}, are analyzed in the remainder of 
the paper. 

The starting Keldysh generating functional reads
\beq
Z^K=\int\mathscr{D}\{a,\bar{a},b,\bar{b},A,\bar{A},B,\bar{B}\}e^{iS^K},
\eeq
where
\beq
S^K=S^K_{sys}+S^K_{res}+S^K_{sys-res}.
\label{Srealt}
\eeq
If we introduce 4-component spinors defined in Nambu-sublattice space for both graphene electrons and leads
electrons
\beq
\phi_{\bf k}(t)\equiv\cvecf{a_{\bf k,\up}(t)}{\bar{a}_{\bf -k,\down}(t)}
{b_{\bf k,\up}(t)}{\bar{b}_{\bf -k,\down}(t)}
\eeq
\beq
\Phi_{\bf k,k_z,\al}(t)\equiv\cvecf{A_{\bf k,k_z,\up,\al}(t)}
{-\bar{A}_{\bf -k,-k_z,\down,\al}(t)}{B_{\bf k,k_z,\up,\al}(t)}{-\bar{B}_
{\bf -k,-k_z,\down,\al}(t)},
\eeq
the actions in Eq.\ref{Srealt} become
\beqm
S^K_{sys}=\int_{\mathcal{C}}dt{1\over N_\tri}\sum_{{\bf k}}
\bar{\phi}_{\bf k}(t)\\\times\square{i\pd_t-\la_{\bf k}\ta^N_z-g_{\bf k}\ta^N_z
\ta^\La_--g^*_{\bf k}\ta^N_z\ta^\La_+}\phi_{\bf k}(t)\\
+U\int_{\mathcal{C}}dt\sum_i\left[\bar{a}_{i,\up}(t)\bar{a}_{i,\down}(t)
a_{i,\down}(t)a_{i,\up}(t)\right.\\\left.+\bar{b}_{i,\up}(t)\bar{b}_{i,\down}(t)
b_{i,\down}(t)b_{i,\up}(t)\right],
\label{sksys}
\end{multline}
\beqm
S^K_{res}=\int_{\mathcal{C}}dt\int{dk_z\over 2\p}\sum_{\al}{1\over N_\tri}
\sum_{{\bf k}}\bar{\Phi}_{\bf k,k_z,\al}(t)\\\times \round{i\pd_t-\e_{{\bf k},k_z}
\ta^N_z}\Phi_{\bf k,k_z,\al}(t),
\end{multline}
and
\beqm
S^K_{sys-res}=\int_{\mathcal{C}}dt\int{dk_z\over 2\p}\sum_{\al}\z_\al{1\over N_\tri}
\sum_{{\bf k}}\\\square{\bar{\Phi}_{\bf k,k_z,\al}(t)\phi_{\bf k}(t)+
\bar{\phi}_{\bf k}(t)\Phi_{\bf k,k_z,\al}(t)}.
\label{ist}
\end{multline}
$\ta^{\nu}_{\pm}$ are 2$\times$2 matrices given by
\beq
\ta^{\nu}_{\pm}={1\over 2}\round{\ta^{\nu}_x\pm i\ta_y^{\nu}},
\eeq
where $\ta^{\nu}_{x,y,z}$ are Pauli matrices. Superscript $\nu$ indicates the space in
which the matrices act; $\La$ ($N$) denotes sublattice (Nambu) space.
The quartic interaction term in Eq.\ref{sksys} is decoupled using Hubbard-Stratonovic fields 
$\De_i^A(t)$ and $\De_i^B(t)$. In the BCS mean-field approximation, where this field is assumed 
static and homogeneous (i.e. $\De_i^A(t)=\De_i^B(t)\equiv\De$), the resulting action of the system 
reads
\beqm
S^K_{sys}=\int_{\mathcal{C}}dt{1\over N_\tri}\sum_{{\bf k}}
\bar{\phi}_{\bf k}(t)\\\times\left[i\pd_t-\la_{\bf k}\ta^N_z-g_{\bf k}\ta^N_z
\ta^\La_--g^*_{\bf k}\ta^N_z\ta^\La_+\right.\\
\left.+U\De\ta_+^N+U\De^*\ta_-^N\right]\phi_{\bf k}(t)
-2U\abs{\De}^2.
\end{multline}
The self-consistency condition for the gap is
\beq
\De=\ang{a_{i,\down}a_{i,\up}}(t)=\ang{b_{i,\down}b_{i,\up}}(t).
\eeq
The time-loop contour integral is carried out by first splitting every field into
two components, labeled "$+$" and "$-$", which reside on the forward and
the backward parts of the time contour, respectively\cite{kamenev,me1,me2}. 
The continuous action then becomes 
\beq
S^K=\int_{-\infty}^\infty dt\square{\mathscr{L}_+(t)-\mathscr{L}_-(t)},
\label{sk}
\eeq
where $\mathscr{L}_\pm (t)$ is the Lagragian corresponding to the action defined in 
Eq.\ref{Srealt} written in terms of $+$ ($-$) fields. When time-ordered products of
Heisenberg fields in the theory are constructed on the Keldysh contour we obtain 
four Green functions
\Beq
iG^T(t,t')&=&\ang{\Y_+(t)\bar{\Y}_+(t')}\nonumber\\
iG^{\tilde{T}}(t,t')&=&\ang{\Y_-(t)\bar{\Y}_-(t')}\nonumber\\
iG^<(t,t')&=&\ang{\Y_+(t)\bar{\Y}_-(t')}\nonumber\\
iG^>(t,t')&=&\ang{\Y_-(t)\bar{\Y}_+(t')}.\nonumber
\Eeq
Because these Green functions are not linearly independent, a linear transformation of the 
fields from the Kadanoff-Baym basis ($+$,$-$) to the Keldysh basis ($cl$,$q$ for bosons; 1,2
for fermions) is commonly performed. For bosons, the barred fields are related to the unbarred
fields simply by complex conjugation and thus the transformation is identical for both, 
\beq
\cvec{\Y_{cl}}{\Y_{q}}={1\over\sqrt{2}}\mat{1}{1}{1}{-1}\cvec{\Y_+}{\Y_-}.
\label{kelrotb}
\eeq
For fermions, unbarred fields transformed in the same manner as Eq.\ref{kelrotb}. For barred
fields, we choose a different transformation\cite{kamenev} 
\beq
\cvec{\bar{\Y}_{1}}{\bar{\Y}_{2}}={1\over\sqrt{2}}\mat{1}{-1}{1}{1}
\cvec{\bar{\Y}_+}{\bar{\Y}_-}.
\eeq
In order to express the Keldysh action, Eq.\ref{sk}, in the Keldysh basis
it is now appropriate to define 8-component spinors for graphene electrons and leads electrons 
defined in the Nambu-sublattice-Keldysh
space. Since we are interested in steady-state properties of the system, it is useful to first Fourier 
transform the fields into frequency space. We define the 8-component spinors as 
\beq
\psi_{k}\equiv\cvece{a^1_{k,\up}}{\bar{a}^1_{-k,\down}}
{b_{k,\up}^1}{\bar{b}^1_{\-k,\down}}{a^2_{k,\up}}
{\bar{a}^2_{-k,\down}}{b^2_{k,\up}}{\bar{b}^2_{-k,\down}}
\quad
\Psi_{k,k_z,\al}\equiv\cvece{A^1_{k,k_z,\up,\al}}
{-\bar{A}^1_{-k,-k_z,\down,\al}}{B^1_{k,k_z,\up,\al}}
{-\bar{B}^1_{-k,-k_z,\down,\al}}{A^2_{k,k_z,\up,\al}}
{-\bar{A}^2_{-k,-k_z,\down,\al}}{B^2_{k,k_z,\up,\al}}
{-\bar{B}^2_{-k,-k_z,\down,\al}},
\eeq
where $k\equiv (\bf k,\w)$ is the energy-momentum 3-vector.
The action (Eq.\ref{sk}) then becomes
\beqm
S^K_{sys}=\int_k\bar{\psi}_{k}\left\{(g^R_0(k)\ta^N_\up-g^R_0(-k)\ta^N_\down)\ta^K_\up
\right.\\(g^A_0(k)\ta^N_\up-g^A_0(-k)\ta^N_\down)\ta^K_\down+g^K_0(k)\ta^N_\up\ta^K_+
\\+g^K_0(k)\ta^N_\down\ta^K_--g_{\bf k}\ta^N_z\ta^\La_--g^*_{\bf k}\ta^N_z\ta^\La_+\\
+\left.U\square{\De_q\ta_+^N+\De_q^*\ta_-^N+(\De_{cl}\ta_+^N+
\De_{cl}^*\ta_-^N)\ta^K_x}\right\}\psi_k\\-2U\square{\De^*_{cl}\De_q+\De^*_q\De_{cl}},
\label{Ssys}
\end{multline}
\beqm
S^K_{res}=\int_k\int{dk_z\over 2\p}\sum_{\al}\bar{\Psi}_{k,k_z,\al}\\
\times\left\{(\tilde{g}^R_\al(k)\ta^N_\up-\tilde{g}^R_\al(-k)\ta^N_\down)\ta^K_\up\right.\\
(\tilde{g}^A_\al(k)\ta^N_\up-\tilde{g}^A_\al(-k)\ta^N_\down)\ta^K_\down\\
\left.+\tilde{g}^K_\al(k)\ta^N_\up\ta^K_++\tilde{g}^K_\al(k)\ta^N_\down\ta^K_-\right\}
\Psi_{k,k_z,\al},
\end{multline}
and
\beqm
S^K_{sys-res}\\=\int_k\int{dk_z\over 2\p}\sum_{\al}\z_\al
\square{\bar{\Psi}_{k,k_z,\al}\psi_{k}+\bar{\psi}_{k}\Psi_{k,k_z,\al}}.
\end{multline}
Here, $\int_k\equiv{1\over N_\tri}\sum_{\bf k}\int{d\w\over 2\p}$, and $\ta_{\up,\down}$ are 
2$\times$2 matrices defined by
\beq
\ta_{\up,\down}=\mat{1}{0}{0}{0},\mat{0}{0}{0}{1}.
\eeq
Superscript $K$ on various $\ta$ matrices indicate that they act in Keldysh space.
$g_0^{R,A,K}(k)$ denote inverse retarded, advanced and 
Keldysh Green functions for non-interacting electrons in the graphene sheet while 
$\tilde{g}_\al^{R,A,K}(k)$ are the corresponding Green functions for lead $\al$. 
For the graphene sheet, they are given by
\Beq
g_0^R(k)&=&\w-\la_{\bf k}+i\de =g_0^{A*}(k)\\
g_0^K(k)&=&2i\de K(\w).
\Eeq
Here, $K(\w)\equiv 1+2n_F(\w)$ where $n_F(\w)$ is the usual Fermi-Dirac distribution function.
$\de$ is an infinitesimal regularization parameter.  For the non-interacting case, 
$g_0^K$ merely serves as a regularization for the Keldysh functional integral.
Because a finite self-energy term is anticipated from the coupling of graphene electrons to the leads 
$g_0^K$ can be safely omitted here (i.e. $g_0^K(k)\approx 0$)\cite{kamenev}.

\subsection{Integrating out the leads}
We now integrate out the leads degrees of freedom in order to obtain an effective theory only in terms
of fields defined on the graphene sheet. The inverse retarded, advanced and Keldysh Green functions
for the leads, $\tilde{g}_\al^{R,A,K}$, are those corresponding to free fermions, and because
the leads are always in thermal and chemical equilibrium the Keldysh Green function is strictly 
related to the retarded and advanced Green functions via the fluctuation-dissipation theorem (FDT). 
They are given by
\Beq
\tilde{g}_\al^R(k)&=&\w-\e_{\bf k,k_z}+i\de =g_\al^{A*}(k)\\
\tilde{g}_\al^K(k)&=&2i\de \tanh\round{\w-\mi_\al\over 2T}.
\Eeq 
Upon integrating over the leads, the resulting self-energy action becomes
\beqm
S_{\S}=\int_k\bar{\psi}_{k}\left\{-\S^R(k)\ta^N_z\ta^K_\up-\S^A(k)\ta^N_z\ta^K_\down\right.
\\\left.-\S^K(k)\ta^N_\up\ta^K_+-\S^K(k)\ta^N_\down\ta^K_-\right\}\psi_k,
\label{Ssigma}
\end{multline}
where
\Beq
\S^R(k)&=&\sum_\al\int{dk_z\over 2\p}\frac{\z_\al^2}{\w-\e_{\bf k}-\e_{k_z}+i\de}
\nonumber\\&=&-i\sum_\al\p\ro t_\al^2=-i\G\\
&=&\S^{A*}(k),\nonumber
\Eeq
and
\Beq
\S^K(k)&=&-2\p i\sum_\al\int{dk_z\over 2\p}\z_\al^2\tanh\round{\frac{\w-\mi_\al}{2T}}
\nonumber\\
&\qquad&\times\de(\w-\e_{\bf k}-\e_{k_z})\nonumber\\
&=&-2i\sum_\al\G_\al\tanh\round{\frac{\w-\mi_\al}{2T}}.
\Eeq
Here, $\G_\al\equiv\p\ro t_\al^2$ measures the effective coupling strength between the layer 
and leads, and $\G=\G_L+\G_R$. $\ro$ is the lead density of states to tunnel into the layer assumed
to be constant. The frequency-independent damping coefficient, $\G$, and the vanishing real energy 
shift that result from our assumptions indicate that the bath is treated as an Ohmic
environment\cite{weiss}. Combining the actions Eqs.\ref{Ssys},\ref{Ssigma}, we obtain the 
dressed inverse Green functions for electrons in the graphene sheet
\Beq
\label{grdress}
g^R(k)&=&\w-\la_{\bf k}+i\G =g^{A*}(k),\\
\label{gkdress}
g^K(k)&=&2i\sum_\al\G_\al\tanh\round{\frac{\w-\mi_\al}{2T}}.
\Eeq
The negative imaginary part of $\S^R(k)$ leads to an irreversible damping in the time-dependent 
Green function $G^R({\bf k},t)$. The damping term formally describes decoherence 
suffered by a propagating electron wave due to incoherent escape and injection of electrons 
into and from the leads. 

At this point, it is convenient to shift the energy scale so that all energies are measured
with respect to $\mi=(\mi_L+\mi_R)/2$. This is equivalent to the following mapping
\Beq
\w&\rightarrow&\w-\mi\nonumber\\
\la_{\bf k}&\rightarrow&\la_{\bf k}-\mi\nonumber\\
\mi_{\al}&\rightarrow&V_\al/2,\nonumber
\Eeq
where $V_{L,R}=\pm V$ and $V\equiv\mi_L-\mi_R$. We assume $V>0$. Following
this choice the inverse retarded Green function, Eq.\ref{grdress}, remains invariant 
while Eq.\ref{gkdress} becomes
\beq
g^K(k)=2i\sum_\al\G_\al\tanh\round{\frac{\w-V_\al/2}{2T}}.
\label{gkdress2}
\eeq
Using the dressed inverse Green functions defined in Eqs.\ref{grdress},\ref{gkdress2}, the
effective action for the graphene sheet is 
\beq
S^{K,eff}_{sys}=\int_{k}\bar{\psi}_k\mathscr{G}^{-1}_{k}\psi_k
-2U\square{\De^*_{cl}\De_q+\De^*_q\De_{cl}},
\label{skefflayer}
\eeq
where the inverse Green function matrix $\mathscr{G}^{-1}_{k}$ is now given by
\begin{widetext}
\beq
\mathscr{G}^{-1}_{k}=\round{\begin{array}{cccccccc}g^R(k)&\De_q&
-g_{\bf k}^*&0&g^K(k)&\De_{cl}&0&0\\\De_q^*&-g^R(-k)&0&g_{\bf k}^*&
\De_{cl}^*&0&0&0\\-g_{\bf k}&0&g^R(k)&\De_q&0&0&g^K(k)&\De_{cl}\\
0&g_{\bf k}&\De_q^*&-g^R(-k)&0&0&\De^{*}_{cl}&0\\0&\De_{cl}&0&0&
g^A(k)&\De_q&-g_{\bf k}^*&0\\\De_{cl}^{*}&g^K(k)&0&0&\De^{*}_q&
-g^A(-k)&0&g_{\bf k}^*\\0&0&0&\De_{cl}&-g_{\bf k}&0&g^A(k)&\De_q\\
0&0&\De_{cl}^{*}&g^K(K)&0&g_{\bf k}&\De_q^{*}&-g^A(-k)\end{array}}.
\label{Ginv}
\end{equation}
\end{widetext}

\subsection{The Mean Field Equations}
\label{mfeqs}
In closed equilbrium, solutions to the mean-field gap and number equations on the honeycomb lattice
have shown that while graphene exhibits a BCS-BEC crossover behaviour away from the Dirac 
point for increasing attractive interaction strength, $u$, superconductivity in graphene at half-filling 
requires a finite attractive interaction\cite{erhai,castro}. In this section, we derive the main 
results of our work which 
are the mean-field gap and number equations in the presence of leads and voltage. Solving these 
equations will allow us to study the effects of dissipation and nonequilibrium current on the gap as a 
function of attractive interaction strength $u$ and filling $n$ and compare the results to the 
equilibrium calculations.
We begin by obtaining an effective theory for the $s$-wave order parameter alone by integrating 
out the graphene electrons. From Eq.\ref{skefflayer}, we obtain
\beqm
iS^{K}_{eff}(\De,\De^*)=Tr\ln[-i\mathscr{G}^{-1}_{k}]\\-2iU\round{\De^{*}_{cl}\De_q+c.c.}.
\end{multline}

\subsubsection{The gap equation}
The self-consistent equation for the gap can be obtained from the following classical 
saddle-point equation

\beq
\label{mfeq1}
\left.{\pd S^K_{eff}\over \pd\De_{q}^{*}}\right|_{\De_{cl}=\De,\De_q=0}=0.
\eeq
In principle, the action may be extremized with respect to $\D_{cl}$ but the corresponding saddle-point
will not be pursued here since it only gives a trivial relation. Eq.\ref{mfeq1} yields,
\Beq
0&=&\left.{\pd S^{K}_{eff}\over\pd\De_q^{*}}\right|_{\De_q=0,\De_{cl}=\De}\nonumber\\
&=&-iTr\left\{\left.{\tau^N_-\over\mathscr{G}^{-1}_{k}}\right|_{\De_q=0,\De_{cl}=\De}
\right\}-\frac{2\De}{U}.
\Eeq
This equation leads to the generalized nonequilibrium gap equation,
\begin{widetext}
\beq
\frac{2\De}{U}=\int_k\frac{4\De\w\sum_\al\G_\al\tanh\round{{\w-V_\al/2\over 2T}}
[((\w+E_{\bf k})^2+\G^2)((\w-E_{\bf k})^2+\G^2)+4\la_{\bf k}^2|g_{\bf k}|^2]}{[(\w-E_+({\bf k}))^2+\G^2]
[(\w-E_-({\bf k}))^2+\G^2][(\w+E_+({\bf k}))^2+\G^2][(\w+E_-({\bf k}))^2+\G^2]}.
\eeq
\end{widetext}
The spectra of the two bands are given by
\beq
E_\pm({\bf k})=\sqrt{\xi^2_\pm({\bf k})+\De^2}\qquad \xi_\pm({\bf k})=
\la_{\bf k}\pm |g_{\bf k}|,
\eeq
and $E_{\bf k}=\sqrt{\la^2_{\bf k}+|g_{\bf k}|^2+\De^2}$. 
After scaling all energies by bandwidth $t$ and evaluating the $\w$-integral we obtain
\beq
{1\over u}={1\over 2\p N}\sum_{\bf k}\square{F_v(\E_+({\bf k}))+F_v(\E_-({\bf k}))},
\label{gapeqTV}
\eeq
where
\[
F_v(x)\equiv{1\over x}\square{\tan^{-1}\round{{v\over 2}+x\over\g}
-\tan^{-1}\round{{v\over 2}-x\over\g}},
\]
and
\[
\E_\pm({\bf k})={E_\pm({\bf k})\over t},\quad u={U\over t},\quad \g_\al={\G_\al\over t},\quad v={V\over t}.
\]
$\g=\g_L+\g_R$ denotes the sum of lead-graphene tunneling rates scaled by $t$.
Eq.\ref{gapeqTV} is the BCS gap equation in the presence of leads ($\g$) and voltage ($v$) and is
the nonequilibrium generalization of Eq.2 in ref.\onlinecite{erhai}. Indeed when one takes the limit as $\g\to 0$
and $v\to 0$ in Eq.\ref{gapeqTV}, the equilibrium gap equation is recovered. 

At low-energies, excitations in graphene at or near half-filling are concentrated near two inequivalent 
Fermi points at the corners of the hexagonal Brillouin zone. In the vicinity of these points, we have
\beq
\la_{\bf k}\approx 3t'-\mi\equiv m\qquad\abs{g_{\pm{\bf K}+{\bf k}}}\approx v_F|{\bf k}|,
\eeq
where $v_F=3t/2$ is the Fermi velocity and $\pm{\bf K}=(\pm 4\p/3\sqrt{3},0)$ are the locations of the 
inequivalent Fermi points. Within this approximation, the quasiparticle dispersions, $\E_\pm({\bf k})$, 
become
\beq
\xi_\pm({\bf k})\approx m\pm \e\qquad\E_\pm({\bf k})\approx \sqrt{\xi_\pm^2+\D^2},
\label{dispctm}
\eeq
where $\e=v_F|{\bf k}|$. Noting that the area per lattice site is $A/N=3\sqrt{3}/4$ the 
conversion from ${\bf k}$-summation to $\e$-integral is given by
\beq
{1\over N}\sum_{\bf k}={3\sqrt{3}\over 4 \p v_F^2}\int_0^D \e d\e,
\eeq
The energy cut-off, set by conserving the total number of states in the Brillouin zone, is 
$D=\sqrt{\sqrt{3}\p}\approx 2.33$ in units of $t$. In the continuum limit, the gap equation then becomes
\beq
{1\over u}={3\sqrt{3}\over 8\p^2v_F^2}\int_0^D \e d\e\square{F_v(\E_+({\bf k}))+F_v(\E_-({\bf k}))}.
\label{neqgapctm}
\end{equation}

\subsubsection{The number density equation}
In equilibrium, the number density is computed using a thermodynamic relation 
$\pd\mathcal{F}_{MF}/\pd\mi =-N_e$. Out of equilibrium, the relation does not hold and the particle density, $n$, 
must be extracted from one of the four Kadanoff-Baym Green functions, $G^<$ using\cite{k-b,mahan}
\beq
n={-i\over 4}\sum_{\s,\La}\int_k G^<_{\s,\La}(k).
\eeq
$\s$ labels the electron spin and $\La\in\{A,B\}$ labels the sublattice in which it propagates. 
In terms of Keldysh Green functions\cite{kamenev},
\beq
n={-i\over 4}\sum_{\s,\La}\int_{k}\square{G^K_{\s,\La}(K)-G^R_{\s,\La}(K)+G^A_{\s,\La}(K)},
\eeq
where $G^{R,A,K}(k)$ are the retarded, advanced and Keldysh Green functions for the graphene electrons. 
These Green functions can be obtained by inverting the matrix, $\mathscr{G}^{-1}(k)$, in Eq.\ref{Ginv}. 
We find that the form of the Green functions is independent of spin and sublattice, and the resulting 
number equation reads
\begin{widetext}
\beq
n={4\g\over N}\sum_{\bf k}\int{d\w\over 2\p}\frac{[1-F(\w,v)](c_6\w^6+c_5\w^5+c_4\w^4
+c_3\w^3+c_2\w^2+c_1\w+c_0)}{[(\w+\E_+)^2+\g^2][(\w+\E_-)^2+\g^2][(\w-\E_+)^2+\g^2]
[(\w-\E_-)^2+\g^2]}.
\label{neqn}
\eeq
$F(\w,v)$ is the zero-temperature nonequilibrium electron distribution, and is given by
\beq
F(\w,v)=\sum_\al{\g_\al\over\g}sgn(\w-v_\al)
={\g_L\over\g}sgn\round{\w-{v\over 2}}+{\g_R\over\g}sgn\round{\w+{v\over 2}}.
\eeq
An exact evaluation of the $\w$-integral in Eq.\ref{neqn} is difficult. However, it can be done in the limit
where the applied bias is assumed small compared to the bandwidth and the dampling coefficient, 
i.e. $v\ll\mbox{min}\{1,\g\}$. Computing the integral up to quadratic order in $v$ the number
density yields 
\beq\bsp
n&={3\sqrt{3}\over 4 \p v_F^2}\int_0^D \e d\e\\&\frac{c_0(10\g^2+\E_+^2+\E_-^2)
+(\g^2+\E_+^2)(\g^2+\E_-^2)[2c_2+c_4(2\g^2+\E_+^2+\E_-^2)+c_6(10\g^4
+6\g^2\E_-^2+\E_-^4+6\g^2\E_+^2+\E_+^4)]}{(\g^2+\E_+^2)(\g^2+\E_-^2)
(16\g^4+\E_+^2(8\g^2+\E_+^2-\E_-^2)+\E_-^2(8\g^2+\E_-^2-\E_+^2))}\\
&-\frac{2}{\p\square{(\E_+^2-\E_-^2)^3+8\g^2\E_+^2(2\g^2+\E^2_+)-8\g^2\E_-^2
(2\g^2+\E^2_-)}}\\
&\times\left\{{\tan^{-1}({\E_+\over\g})\over\E_+}\left[c_1(\E_+^2-\E_-^2-4\g^2)
+c_3(\E_+^4+\g^2\E_-^2+4\g^4-\E_+^2(\E_-^2-3\g^2))\right.\right.\\
&\left.+c_5[\E_+^6+6\g^2\E_+^4+9\g^4\E_+^2-\E_-^2(\g^4-6\g^2\E_+^2+\E_+^4)-4\g^6]\right]
-(-\leftrightarrow+)\\
&\left.+\g\ln\round{\frac{\g^2+\E_+^2}{\g^2+\E_-^2}}\round{2c_1+c_3(\E_+^2+\E_-^2+2\g^2)
+2c_5(\E_+^2\E_-^2-\g^2\E_+^2-\g^2\E_-^2-3\g^4)}\right\}\\
&+(2x-1)\frac{2\g c_0}{\p(\g^2+\E_+^2)^2(\g^2+\E_-^2)^2}v
+\frac{\g c_1}{2\p(\g^2+\E_+^2)^2(\g^2+\E_-^2)^2}v^2,
\label{neqnctm}
\end{split}\eeq
where $x=\g_L/\g$ and $\E_\pm$ are given by Eq.\ref{dispctm}. The coefficients $c_0,\dots,c_6$
are dependent on $\E_\pm$, $\xi_\pm$ and $\g$, and are defined as
\Beq
c_6&=&1,\nonumber\\
c_5&=&\xi_++\xi_-,\nonumber\\
c_4&=&3\g^2-\frac{\E^2_++\E^2_-}{2},\nonumber\\
c_3&=&2[\xi_+(\g^2-\E^2_-)+\xi_-(\g^2-\E^2_+)],\\
c_2&=&3\g^4+(\E^2_+-\E^2_-)^2+\g^2(\E_-^2+\E_+^2)-{\E_-^4\over 2}-{\E_+^4\over 2},\nonumber\\
c_1&=&\xi_-(\g^4+2\g^2\E^2_++\E^4_+)+\xi_+(\g^4+2\g^2\E^2_-+\E^4_-),\nonumber\\
c_0&=&{1\over 2}(\E_-^2+\g^2)(\E_+^2+\g^2)(\E_+^2+\E_-^2+2\g^2).\nonumber
\Eeq
\end{widetext}
It can be easily verified that in the limit of $\g\to 0$ and $v\to 0$, Eq.\ref{neqnctm} reduces to the equilibrium 
number equation (c.f. Eq.3 in ref.\onlinecite{erhai}). The mean-field equations Eqs.\ref{neqgapctm},\ref{neqnctm} 
are the central results of this work. These equations will be analysed in the remainder of the paper.

\section{Results}
\label{results}
Our main focus will be on obtaining and analyzing gap phase diagrams in the parameter space of 
interaction strength ($u$) and number density ($n$) for various leads-graphene couplings 
($\g_L$,$\g_R$) and external biases ($v$). A previous work on closed equilibrium graphene\cite{erhai} 
revealed that, at half-filling, the superconducting instability of the semi-metallic phase requires a critical 
attractive interaction strength $u_c$ and, thus, the gap vanishes up to $u_c$. 
Away from half-filling, the metallic phase is immediately unstable to superconductivity for arbitrarily
weak attractive interaction strength. As a result, the gap remains finite for any finite $u$ and the system displays
a typical BCS-BEC crossover behaviour. In this section we quantitatively discuss the effects of dissipation 
and nonequilibrium current on the gap phase diagram by numerically solving the generalized 
mean-field equations, Eqs.\ref{neqgapctm},\ref{neqnctm}.
The following sections will show that a dramatic modification to the phase diagram is 
observed by the mere coupling of graphene to its environment, even in the absence of nonequilibrium 
current. We find that the effects of external biases in addition to dissipation does not substantially 
alter the qualitative features of the phase diagram from the case in which the system is subject to dissipation 
alone. However, as the following sections will discuss, the application of an external bias leads to shifts in the 
metallic region surrounding half-filling which result from voltage-induced changes in the graphene electron 
density. The results presented here are applicable to the case of small biases ($v\ll\mbox{min}\{1,\g\}$); 
effects of large biases are not considered here.

\subsection{Finite lead-layer coupling $\g\ne 0$, zero voltage ($v=0$)}
\label{results0v}
First, we begin with the case in which the lead-graphene-lead heterostructure is in thermodynamic equilibrium. 
In particular, this is the situation where $\mi_L=\mi_R=\mi_{res}$, and in the long-time limit 
$\mi_{sys}=\mi_{res}$ is maintained. Here, electron tunneling processes between the central graphene system 
and the leads is providing a mechanism for decoherence for the particles in the system ($\g\ne 0$), but an 
external bias that explicitly breaks time-reversal symmetry of the heterostructure is absent ($v=0$). Consider 
the case where the central graphene sheet is in a superconducting phase. Because of its coupling to the
leads one can envisage a situation in which an electron that constitutes 
a Cooper pair escapes into the leads. Because the leads are assumed to be infinite the electron that has escaped 
the system is completely lost in the leads and as a consequence looses its coherence with its former partner. 
Although a different electron may enter the system from a lead within a time-scale of $\ta_{tun}\sim 1/\G$, the
electron will not necessarily pair with the widowed electron since it completely lacks coherence to do so. 
Because dissipation effectively acts as a pair-breaking mechanism we expect a suppression of the 
gap throughout the entire region of the phase diagram.

Fig.\ref{fig:gap} plots the gap phase diagrams for various leads-graphene coupling strengths ($\g$). 
Fig.\ref{fig:gap}(a) corresponds to the closed equilibrium case which has been obtained 
previously\cite{erhai}. Fig.\ref{fig:gap}(b),(c) display the behaviour of the gap as $\g$ 
is increased. It is apparent from these plots that the suppressed region in the gap (dark blue region)
grows as $\g$ is strengthened. Regions of large gap values corresponding to the region with large 
$u$ also displays an overall suppression in the gap as $\g$ is increased. The qualitative
features of the diagrams are consistent with the expectation described above. Let us now discuss 
the results more quantitatively.

\subsubsection{Half-filling ($n=1$)}
\begin{figure}[h]
\includegraphics[scale=0.5,angle=270]{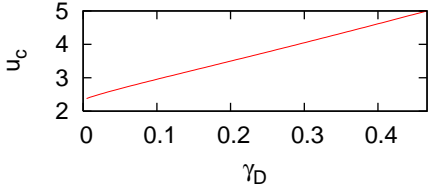}
\caption{\label{fig:ucvsg} The plot of critical coupling $u_c$ as a function of reduced 
leads-graphene coupling $\g_D=\g/D$.}
\end{figure}

\begin{figure}[h]
\includegraphics[scale=0.5,angle=0]{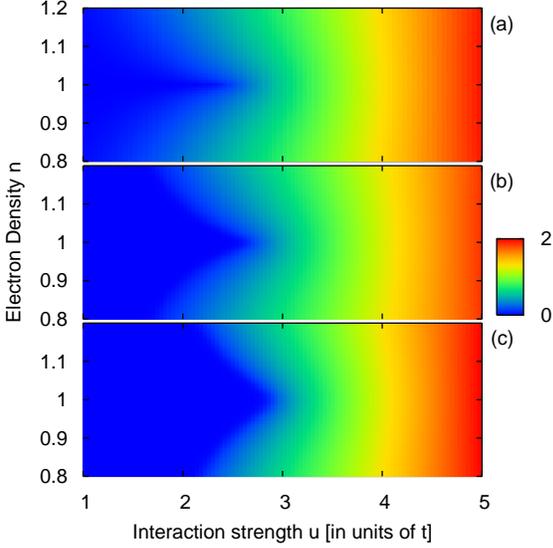}
\caption{\label{fig:gap} Plots of the BCS gap, $\De$, in the parameter space of attractive 
interaction strength $u$ and electron density $n$. The three diagrams correspond to different values
of leads-graphene coupling strengths. In (a), the system is closed, i.e. $\g=0$, while in (b) and (c)
$\g=0.1$ and $\g=0.2$, respectively. As the coupling is increased, the blue region in the phase diagram,
where the gap is small, grows. In parts of the blue regions in (b) and (c) the gap is zero even for $n\ne 1$,
indicating that a metal-superconductor quantum phase transition emerges in the presence of dissipation.}
\end{figure}

For the closed equilibrium case at half-filling ($\g=v=0$ and $n=1$) the semimetal-superconductor 
transition is possible mainly because the divergent nature of the integral on the right hand side of
Eq.\ref{neqgapctm} is cured by particle-hole symmetry. When the
integral is convergent, it is clear that a solution to the gap equation does not exist for small $u$
where $u^{-1}$ becomes larger than the integral. The value of the critical interaction parameter 
at which the transition occurs can be easily quantified. At half-filling the number equation, 
Eq.\ref{neqnctm}, is satisfied by $m=3t'-\mi=0$, and thus at the critical point ($n=1$, $\De=0$, and $m=0$) 
the gap equation reads
\beq
{1\over u_c}={3\sqrt{3}\over 4 \p v_F^2}\int_0^D d\e={1\over 2.33}.
\eeq
For any $u<u_c$ the equations cannot be solved with any real $\De$ and the system enters the 
semimetallic phase. In the presence of dissipation ($\g>0$) the number equation is still solved
by $m=0$ at half-filling, and the gap equation at the critical point yields
\Beq
{1\over u_c}&=&{3\sqrt{3}\over 2\p^2v_F^2}\int_0^Dd\e\tan^{-1}\round{\e\over\g}
\nonumber\\
&=&{3\sqrt{3}D\over 2\p^2v_F^2}\square{\tan^{-1}(\g_D^{-1})-
{\g_D\over 2}\ln\round{1+\g_D^{-2}}},
\label{ucneq}
\Eeq
where the reduced coupling strength is given by $\g_D=\g/D$. The integral on the right hand side
of Eq.\ref{ucneq} is convergent and, thus, tells us that the semimetal-superconductor transition 
exists in the presence of dissipation at half-filling. The behaviour of $u_c$ as a function of $\g_D$ 
is plotted in Fig.\ref{fig:ucvsg}.  We see that the value of $u_c$ increases as $\g$ is increased. 
This is consistent with the above considerations from which we expect that a larger interaction 
parameter is necessary to achieve pairing since leads-induced decoherence generally suppresses 
superconductivity. The phenomenon can also be observed in Fig.\ref{fig:gap} where the apex of 
the blue region shifts right for larger $\g$. The plots show that at $\g=0$ $u_c$ converges to the 
closed equilibrium value of $u_c\sim 2.33$ as predicted by previous calculations. 

\subsubsection{Away from half-filling ($n\ne 1$)}
In the closed equilibrium case away from half-filling, $m\ne 0$ and the critical point condition
becomes 
\beq
{1\over u_c}={3\sqrt{3}\over 4 \p v_F^2}\int_0^D \e d\e
\square{{1\over\abs{m+\e}}+{1\over\abs{m-\e}}}=\infty.
\eeq
The divergence of the integral results in a solution with $\De>0$ for any small $u>0$. 
This gives $u_c=0$ implying that Cooper instability occurs for any finite $u$ away from 
half-filling. Let us now investigate how this is modified when $\g$ is finite.

What is notable in Fig.\ref{fig:gap} is the expansion of the blue region, where the gap is 
small, as $\g$ is increased. The question is whether or not the typical BCS-BEC crossover 
behaviour observed in the closed equilibrium case is a correct physical picture away from
half-filling for finite $\g$. The external baths acting as a pair-breaking mechanism makes 
the issue subtle. The pair-breaking perturbation in a superconductor with magnetic impurities
has been shown\cite{abrikosov,matthias} to strongly suppress the transition temperature of 
the superconductor. Therefore, when such perturbation is strong enough the gap may vanish 
completely and give rise to a metal-superconductor quantum phase transition at finite 
doping. The question of whether or not the gap vanishes away from half-filling 
depends on the convergence of the integral in the gap equation. At $v=0$, the 
generalized gap equation becomes
\beq
{1\over u}\propto\int_0^D \e d\e\square{{1\over \E_+}\tan^{-1}\round{{\E_+\over\g}}+(+\to -)}.
\label{gapeqngen}
\eeq
\begin{figure}[h]
\includegraphics[scale=0.5,angle=0]{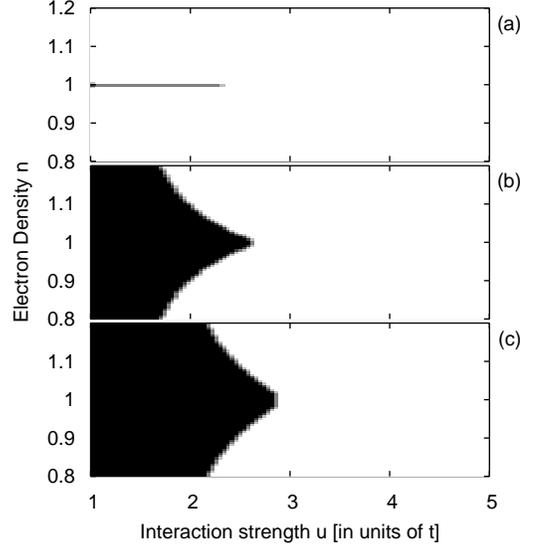}
\caption{\label{fig:ucs} The dark areas above show regions in the phase diagram where the gap
equation lacks a solution for any finite $\De$; the gap vanishes in these regions. As in Fig.\ref{fig:gap},
the system is closed for plot (a) while $\g=0.1$ and $\g=0.2$ in plots (b) and (c), respectively.}
\end{figure}
We see that for any $m$ (i.e. regardless of being at half-filling or not), the integral is convergent because for 
any small $\E_\pm$, which is the source of divergence, the arctan factor nullifies the divergence. This 
implies a finite $u_c$ at both half-filling and away from half-filling. Consequently, the system should undergo a
superconductor-to-metal phase transition as the interaction parameter is lowered. Notice that the analysis 
above infers that the system will eventually enter the metallic phase as $u$ is decreased for any density. 

Fig.\ref{fig:ucs} explicitly shows regions in the gap phase diagram where the gap equation lacks a
solution with any positive $\De$. The diagrams are plotted for the same values of $\g$ as in 
Fig.\ref{fig:gap}. The black regions are where the gap equation is solutionless and represents
a (semi)metallic phase. Clearly, as $\g$ is increased, the metallic region expands. 
We find that the superconducting (white) and metallic (black) regions are separated by a second-order 
phase transition. 

The fact that the central graphene sheet becomes metallic away from half-filling certainly defies expectations
based on the intrinsic properties of graphene. For closed graphene the 
semimetal-superconductor quantum criticality emerges at half-filling because of the absence of electron-electron
screening that results from the vanishing density of states at the Dirac point. As stated earlier in this section, 
the physics of the metallic region can be understood using the phenomenon of leads-induced decoherence, 
and, thus, is not described by intrinsic properties of the central graphene sheet.

\subsection{Effect of voltage, $v\ne 0$}
\label{resultsfv}
So far, we have discussed the effect of leads-induced dissipation on the gap phase diagram 
in the absence of voltage. We now consider the effects of driving an out-of-plane 
charge current though the superconducting graphene sheet. Here, we are limited to the regime 
of small voltages, specifically $v\ll\mbox{min}\{1,\g\}$. As mentioned before, we assume 
$v\propto\mi_L-\mi_R>0$ and allow the relative strengths of the two couplings to the 
leads, $\g_L$ and $\g_R$, to vary. In the absence of current ($v=0$), the gap equations 
depends only on the sum of these couplings $\g=\g_L+\g_R$. But Eq.\ref{neqnctm} 
shows that in the presence of current ($v\ne 0$) the number density now depends on these couplings
independently and depending on the relative strengths of these couplings the dominant correction term
may change sign. The main qualitative modifications to the gap phase diagram in the presence of 
finite voltage reflects the influence of this correction term. 

\begin{figure}[h]
\includegraphics[scale=1,angle=0]{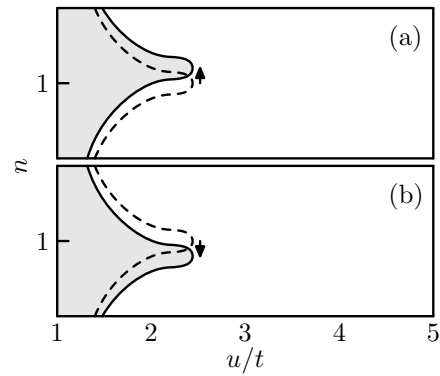}
\caption{\label{fig:gapdiff} A cartoon plot showing the effect of voltage on the 
boundary of the metallic region. The dashed lines in both plots denote the boundary
at $v=0$. The shaded area is the metallic region after a steady-state bias is applied.
In both plots, the applied boltage is $v=0.1$. However, $\g_L>\g_R$ in (a) while
$\g_L<\g_R$ in (b). Essentially, the voltage-induced modification is to shift the 
metallic region to higher values in density or to lower values depending on the
polarity of the voltage and the lead-coupling asymmetry.}
\end{figure}

In the small voltage regime and for $\g<1$, the dominant correction term gives a correction of
order $\g v\ll 1$ to the number density, which is of order unity.  Because the modifications
to the gap phase diagram due to voltage is expected to be small the effect is more clearly
seen by plotting the difference in gap values at finite and zero voltage. This is shown in 
Fig.\ref{fig:gapdiff}.
The gap difference is plotted here for $2v=\g=0.2$ in the vicinity of the apex region. 
In Fig.\ref{fig:gapdiff}(a) $\g_L/\g_R=4$ while in Fig.\ref{fig:gapdiff}(b) $\g_L/\g_R=0.25$.  
The plots reveal that the metallic region (blue region in Fig.\ref{fig:gap}) shifts vertically  
away from half-filling. The figure shows that for $\g_L/\g_R=4$ the apex shifts up while 
for $\g_L/\g_R=0.25$ it shifts down. 
Given $\mi=(\mi_L+\mi_R)/2$ and $v>0$ the lowest order voltage correction in Eq.\ref{neqnctm} 
tells us that the number density increases or decreases depending on the asymmetry of the lead couplings.
If $\g_L>\g_R$, $n$ increases, while if $\g_L<\g_R$, $n$ decreases. The gap equation yields
the largest value of $u_c$ given $\g$ and $v$ when $m=0$. Thus, the above observation tells us that, 
for $\g_L>\g_R$, $m=0$ is achieved not at half-filling as in the equilibrium case but at $n>1$. This shifts the 
apex upward. The opposite occurs for $\g_L<\g_R$. The nonequilibrium gap equation is convergent for all $\mi$, 
thus, a metallic phase is once again expected at all densities.

\section{Conclusion}
\label{conclusion}
\begin{figure}[b]
\begin{center}
\includegraphics[scale=0.8,angle=0]{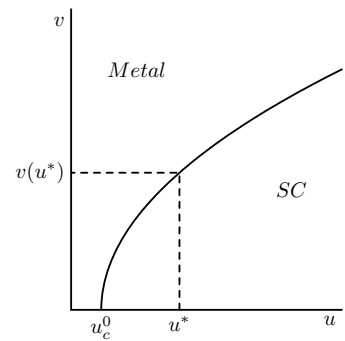}
\caption{\label{fig:vvsu} A plot of $u_c$ vs. $v$ for a fixed $\mi$. The plot line
separates the metallic and superconducting phases of our system. Adjusting
$\mi$ will tune the location of $u_c^0$ on the $x$-axis but the general shape of the curve
is not modified.}
\end{center}
\end{figure}
In conclusion, we have theoretically studied the effects of dissipation and nonequilibrium 
drive on the 
properties of superconducting graphene. An external steady-state current was perpendicularly 
driven through the graphene sheet by attaching it to two leads which were
equilibrated at two constant, but different, chemical potentials. The mean-field
BCS theory of superconductivity on graphene was extended to the nonequilibrium
situation by formulating the theory on the Keldysh contour. After obtaining nonequilibrium
gap and number density equations we studied the BCS gap as a function of attractive 
interaction strength $u$ and electron density $n$ for various lead-graphene 
coupling strengths $\g$ and voltages $v$. We have shown that dissipation results in a 
suppression of the BCS gap at both zero and finite voltages. We argued that the coupling
of the graphene sheet to external baths acts as a pair-breaking mechanism because it causes 
an electron that constitutes a Cooper pair to escape into the leads. Once an electron leaves the 
scattering region, it looses coherence with its time-reversed partner and the destruction of
the Cooper pair entails.

A quantitative understanding of why the gap is significantly suppressed by dissipation can be 
gained by observing how dissipation affects the gap equation. Recall that the BCS gap
equation for an ordinary superconductor\cite{note1} in closed equilibrium is given by
\beq
\De=uTN(0)\sum_n\frac{\De}{\sqrt{\w_n^2+\De^2}}.
\label{gengapeqn}
\eeq
$N(0)$ is the density of states at the Fermi energy, and $u>0$ is the attractive interaction strength.
A general result for these ordinary superconductors is that the gap equation (Eq.\ref{gengapeqn}),
and hence the gap, is unaffected by time-reversal-invariant perturbations. Take, for example, the
influence of non-magnetic impurities on the superconducting state. The gap equation obtained
after invoking disorder-averaging and the Born approximation reads
\beq
\De=uT\tilde{N}(0)\frac{\tilde{\De}}{\sqrt{\tilde{\w}_n^2+\tilde{\De}^2}},
\label{gengapeqntrs}
\eeq
where $\tilde{\w}$ and $\tilde{\De}$ are frequency and order parameter renormalized by the
perturbation\cite{maki,crisan,adg}, and $\tilde{N}(0)$ is the density of states in the presence of the 
perturbation. The essential point is that $\tilde{\w}$ and $\tilde{\De}$ are related to their unrenormalized 
counterparts by a common factor $\eta=\eta(\w_n,\De)$, i.e.
\Beq
\tilde{\w}&=&\eta\w,\nonumber\\
\tilde{\De}&=&\eta\De.\nonumber
\Eeq
Because this factor $\eta$ cancels out in Eq.\ref{gengapeqntrs},  
the gap equation remains invariant and leads to the result that the gap is unaffected by 
non-magnetic impurities\cite{ag}. 

Imagine now that a pure ordinary superconductor is coupled to an external bath in equilibrium. 
The Nambu-Gorkov equations can be straightforwardly derived for this case,
\Beq
\label{ngeqn1}
& &\round{i\w_n+isgn(\w_n)\G-\xi_{\bf k}}G+\De F=1\\
\label{ngeqn2}
& &\round{i\w_n+isgn(\w_n)\G+\xi_{\bf k}}F+\De G=0.
\Eeq
where the ordinary and anomalous Green functions are given by
\Beq
G({\bf k},\w_n)&=&-\int_0^\be d\ta\ang{T_\ta c_{{\bf k},\up}(\ta)c^\dag_{{\bf k},\up}(0)}e^{i\w_n\ta}
\nonumber\\
F({\bf k},\w_n)&=&-\int_0^\be d\ta\ang{T_\ta c_{{\bf k},\up}(\ta)c_{-{\bf k},\down}(0)}e^{i\w_n\ta}.
\nonumber\\\nonumber
\Eeq
We immediately see from Eqs.\ref{ngeqn1},\ref{ngeqn2} that $\w$ and $\De$ scale asymmetrically,
namely, 
\beq
\tilde{\w}=\eta\w\quad \tilde{\De}=\De;\qquad\eta=1+{\G\over\abs{\w_n}}.
\label{asymm}
\eeq
Here, $\G$ is the rate at which electrons decay into the bath. The asymmetry in the renormalization 
of $\w$ and $\De$ (Eq.\ref{asymm}) greatly affects the gap equation, Eq.\ref{gengapeqntrs}, and 
shows how dissipation can affect the gap significantly. This is consistent with the qualitative
argument given above.

The emergence of the metal-superconductor quantum phase transition in the graphene
subsystem at both zero and finite voltages gives rise to the possibility of inducing the phase 
transition using external bias. While fixing the average chemical 
potential $\mi$ to some value, $v$ can be changed
by adjusting $\mi_L$ and $\mi_R$ symmetrically about $\mi$. $u_c$ is obtained from the gap 
equation in this situation by fixing $\D=0$ and $\mi$ to some value. Fig.\ref{fig:vvsu} shows
a generic plot of $u_c$ as a function of voltage.
If the interaction strength,$u$, of the system is at $u=u^*$, then for $v<v(u^*)$ the system will be 
metallic. However, when $v$ is increased and passes $v=v(u^*)$, the system will become
superconducting. $u_c^0$ can be tuned by adjusting the average chemical potential $\mi$. 
It is clear from Eq.\ref{neqnctm} that when the average chemical potential $\mi$ is fixed,
the electron density can change as a function of voltage.

\textbf{Acknowledgment:} The authors would like to thank Michael Lawler, Eun-Ah Kim, 
Erhai Zhao, Arun Paramekanti, and Ilya Vekhter for helpful discussions. This research 
was supported by NSERC of Canada (S.T.), The Canada Research Chair program, Canadian 
Institute for Advanced Research, and KRF-2005-070-C00044 (Y.B.K.).



\begin{thebibliography}{99}
\bibitem{note1} Here, we are considering an ordinary Fermi liquid (with quadratic dispersion relation)
in the BCS superconducting phase.
\end{thebibliography}
\end{document}